\documentclass[aps, prx, reprint,10pt,longbibliography,superscriptaddress]{revtex4-2}

\usepackage{upgreek,mathtools}
\usepackage{color}
\usepackage{graphicx}
\usepackage{nccmath}
\usepackage{bm}
\usepackage{hyperref}
\hypersetup{colorlinks,breaklinks,
            urlcolor=[rgb]{0,0,0.36},
            linkcolor=[rgb]{0,0,0.36},
            citecolor=[rgb]{0,0,0.36}}
\usepackage[mathlines]{lineno}
\usepackage{dcolumn}
\usepackage{mathtools}  
\usepackage{dsfont}
\usepackage{comment}
\usepackage{physics}
\usepackage{xcolor}
\usepackage{adjustbox}

\newcommand{\Harvard}{Department of Physics, Harvard University, Cambridge, Massachusetts 02138, USA.}

\newcommand{\MaxP}{Max Planck Institute for the Structure and Dynamics of Matter, Luruper Chaussee 149, 22761 Hamburg, Germany.}

\newcommand{\BNL}{Condensed Matter Physics and Materials Science Division, Brookhaven National Laboratory, Upton, New York 11973, USA.}

\newcommand{\Oxford}{Clarendon Laboratory, University of Oxford, Parks Road, Oxford OX1 3PU, UK.}

\newcommand{\ETH}{Institute for Theoretical Physics, ETH Zurich, 8093 Zurich, Switzerland.}

\begin{document}

\title{Gap Inhomogeneity in Cuprates: a view from Two-Dimensional Josephson Echo Spectroscopy}

\author{Alex~G\'{o}mez~Salvador}
\affiliation{\ETH}

\author{Ivan~Morera}
\affiliation{\ETH}

\author{Marios~H.~Michael}

\affiliation{\MaxP}

\author{Pavel~E.~Dolgirev}

\affiliation{\Harvard}

\author{Danica~Pavicevic}
\affiliation{\MaxP}

\author{Albert~Liu}

\affiliation{\BNL}


\author{Andrea~Cavalleri}
\affiliation{\MaxP}
\affiliation{\Oxford}

\author{Eugene~Demler}
\affiliation{\ETH}
\date{\today}

\begin{abstract}
    Novel theoretical developments have allowed to connect microscopic disorder in bosonic collective excitations to the signatures in two-dimensional terahertz spectroscopy~\cite{salvador2025Echo_theory}.
    Here, we employ this framework to analyze the recently measured Josepshon echoes in optimally doped La$_{2-x}$Sr$_x$CuO$_4$ in Ref.~\cite{Liu_2023_echo}. 
    We consider the spatial gap inhomogeneities ---observed in scanning tunneling microscopy--- as input for the disorder in the superfluid density, and compute the resulting echo peaks.
    The excellent agreement supports the interpretation that the gap inhomogeneity arises solely from pairing gap fluctuations, with no evidence for non-superconducting competing local orders.
    Finally, we study the microscopic origin of the inelastic processes, contributing to the damping of the Josephson plasmon at low temperatures, and conclude that it can be attributed to nodal quasiparticles.
\end{abstract}

\maketitle

\textit{Introduction.---} Understanding the nature of competing orders in high-temperature superconductors remains one of the central challenges in condensed matter physics. A key observation comes from scanning tunneling microscopy (STM), which reveals the existence of two distinct electronic energy gaps~\cite{Norman_1995_phenomenological,Renner_STM_review,Ronen_2016_charge,yazdani2016spectroscopic,Gozlinski_2024} with a pronounced spatial inhomogeneity~\cite{Pan2001,kato2005inhomogeneous}. 
Are both gaps manifestations of superconducting correlations, or does one of them arise from a competing local order parameter, such as a charge- or spin-density wave~\cite{RevModPhys.75.1201,rev_Xray_competing_orders,charge_correl_cuprates,Torre_2015}, or even more exotic states involving spin liquids~\cite{RVB_SL_cuprates_2006,bonetti2025criticalquantumliquidscuprate}?
Traditional STM experiments are limited in their ability to resolve this issue, as they cannot discriminate between quasiparticle gaps originating from different electronic orders. Notable exceptions are Josephson scanning tunneling spectroscopy~\cite{Graham_2019_Josephson_STM} and recent STM noise measurements~\cite{niu2024equivalencepseudogappairingenergy}, strongly hinting towards the preformed pairs hypothesis, and therefore the absence of non-superconducting local orders. In this paper, we confirm two-dimensional terahertz spectroscopy (2DTS) of Josephson plasmons as a novel optical route to address these questions, providing a theoretical link between the dynamical response of the spatially inhomogeneous gap landscape in cuprates and physics inaccessible to both traditional STM and linear reflectivity measurements.

    Below the superconducting transition temperature $T_c$, the CuO planes of high-temperature cuprate superconductors develop long-range order. This gives rise to coherent interlayer charge transfer via Josephson tunneling, which manifests as a sharp plasma edge in the reflectivity signal at the Josephson plasma resonance (JPR)~\cite{laplace2016josephson,Sellati2023GeneralizedJPs}. The JPR has been extensively investigated for its appeal as an optical probe of both interlayer coherence and in-plane superconductivity~\cite{VANDERMAREL19911,Dordevic_2003,Tamasaku_1992,2023_THG_Averitt}. Even at low temperatures $T\ll T_c$, the plasma edge presents a sizable broadening, which increases as temperature grows and becomes unresolvable as $T$ approaches $T_c$. Disorder, nodal quasiparticles, phase fluctuations, or competing orders are among the possible mechanisms to explain the origin of the JPR broadening~\cite{Basov_review_2005,LaForge_2010_possibility}. Identifying and characterizing the origin of the plasmon linewidth and its temperature dependence is thus fundamental to understand the nature of cuprate superconductivity. 
However, linear reflectivity, which probes only two-point correlators—specifically $\expval{j_zj_z}$ for the JPR—cannot disentangle different scattering processes, as they all contribute indistinguishably to the plasma-edge linewidth.

The limitations of linear spectroscopy can be overcome by employing more sophisticated spectroscopic techniques capable of resolving higher order correlation functions~\cite{Mukamel}. Of special interest is the terahertz analogue of multidimensional optical spectroscopy~\cite{hamm_zanni_2011}, known as two-dimensional terahertz spectroscopy (2DTS)~\cite{Liu2025Qmat,huang2025review}. Recently, 2DTS has been employed to study a wide range of condensed-matter systems, such as superconductors~\cite{Novelli_persistent_2020, Zhang2023revealing,Sijie_2023,puviani2022quench,puviani2023quench,katsumi_revealing_2024_benfatto,puviani2024theory,salvador2024principles,katsumi2024amplitudemodemultigapsuperconductor,tsuji2025twodimensionalcoherentspectroscopydisordered}, correlated electrons~\cite{barbalas2023energy,chen2024multidimensionalcoherentspectroscopycorrelated}, and spin systems~\cite{Lu2017,Lin2022,zhang2024terahertz1,zhang2024terahertz2,parameswaran2020asymptotically1,Blank2023,wan2019resolving,liebman2023multiphotonspectroscopydynamicalaxion,Choi2020SpinLiquid,mcginley2022signatures,McGinley2024Anomalous,potts2024signaturesspinondynamicsphase}. 
In particular, measurements in optimally doped La$_{2-x}$Sr$_x$CuO$_4$ (LSCO) have shown that the rephasing signal of 2D terahertz spectroscopy is capable of disentangling the presence of elastic and inelastic scattering processes in the broadening of the JPR~\cite{Liu_2023_echo}, paving the way for the characterization of its different broadening mechanisms. 

In this Letter we apply the theoretical framework for 2DTS of bosonic collective excitations developed in Ref.~\cite{salvador2025Echo_theory} to analyze the experimental measurements of Josephson echoes reported in Ref.~\cite{Liu_2023_echo}. Leveraging the unique capabilities of the echo peak, we are able to connect quantitatively the spatial inhomogeneity in the superfluid density with the spectroscopic signatures of disorder. Furthermore, we provide a microscopical description of the inelastic processes contributing to the broadening of the JPR at low temperatures.
\vspace{0.25cm}

\textit{The model.---} We describe the nonlinear electromagnetic response of layered superconductors following Refs.~\cite{PhysRevB.50.12831,PhysRevB.50.12831,koshelev1999fluctuation,Basov_review_2005}. The layered superconductivity is described by means of the Lawrence-Doniach model~\cite{Lawrence_Doniach_1970}. For probe frequencies $\omega_p\ll2\Delta$ (and $T\ll\Delta$), amplitude fluctuations can be neglected and the dynamics are fully encoded in the order parameter phase $\varphi_n(\boldsymbol{r},t)$, with $n$ the layer index. 
After incorporating electrodynamics, in the clean limit, and in the absence of normal fluid transport, the equation of motion for the Gauge invariant phase difference
\begin{equation}
    \psi_n = \varphi_n - \varphi_{n + 1}  - \frac{2\pi}{\Phi_0} \int_{n s}^{(n + 1) s} dz\, A_z
\end{equation} 
reads
\begin{align}
    \partial_t^2 \psi_n - \nabla^2 L_{nm}\psi_m + \omega_{\text{JP}}^2 \sin\psi_n = 0,
    \label{eqn: Bulaevskii}
\end{align} 
with $\Phi_0 = 2\pi c/(2 e)$, $A_z$ the $z$-component of the vector potential, $\omega_{\text{JP}}$ the Josephson plasma resonance, and
\begin{align}
    L_{nm} = \frac{c^2s^2}{\epsilon_\infty\lambda_{ab}^2N} \sum_{k_z} 
    \frac{e^{ik_z(n - m)}}{ 2(1 - \cos k_z) + s^2/\lambda_{ab}^2}.
    \label{eqn: def L}
\end{align}
Here $N$ is the total number of layers ($ k_z = 2\pi n/N$, with $n$ an integer and $0<n<N-1$), $c$ the speed of light, $s$ the c-axis interlayer distance, $\lambda_{ab}$ the London penetration length, and $\epsilon_\infty$ the high frequency dielectric constant. We note that it is possible to obtain the same phase dynamics from an action-based method~\cite{Sellati2023GeneralizedJPs}.
Apart from the clean dynamics of $\psi_n$, we consider the possibility of scattering events that result in a finite excitation lifetime. We divide the sources of broadening for the JPR into two classes: (i) elastic scattering processes, originating from static disorder, which correspond to energy-conserving scattering events; and (ii) inelastic scattering processes, involving energy transfer, such as quasiparticle damping.

First, we consider elastic scattering events for the Josephson plasmons arising from the spatial inhomogeneity of the superfluid density, proportional to $\omega_{\text{JP}}^2$. It is well known that layered superconductors present a spatial inhomogeneity of the single particle energy gap at the nanometer scale~\cite{Pan2001,kato2005inhomogeneous}. The origin of said electronic energy gap is a topic of current debate, and explanations involving competing orders beyond superconducting correlations have been put forward. In this work, we assume that the entire gap, together with its spatial variations, is of superconducting origin—an assumption supported by our results. 
To capture the disorder effects in our modeling, we introduce a source of elastic scattering by promoting $\omega_{\text{JP}}^2\rightarrow\omega_{\text{JP}}^2(1+V_n(\boldsymbol{r}))$ in Eq.~\eqref{eqn: Bulaevskii}. 
The static random potential $V_n(\boldsymbol{r})$ is assumed to be mean-free, Gaussian, and delta-correlated, i.e., $\overline{V_n(\boldsymbol{r})} = 0$ and $\overline{V_n(\boldsymbol{r})V_m(\boldsymbol{r}')}=2\pi\overline{V}^2\xi^2\delta_{nm}\delta^{(2)}(\boldsymbol{r}-\boldsymbol{r}')$. The delta-correlated assumption is justified due to the big mismatch between the characteristic lengthscales of the plasmon $\omega_{\text{JP}}/c\sim100\,\mu$m and the disorder $\xi\sim10$ nm, such that $\xi\ll\omega_{\text{JP}}/c$. The spatial inhomogeneity of the STM gap directly influences the inhomogeneity of the JPR, as $\omega_{\text{JP}}^2\propto\Delta^2$. Note that this relation holds only under the current assumption that the STM gap arises entirely from superconducting pairing. From STM measurements of LSCO~\cite{kato2005inhomogeneous}, we estimate a mean gap of $\bar{\Delta} = 10$ meV, with a relative variance of $\sigma_\Delta/\bar{\Delta} = 0.19$, and a correlation length of $\xi=10$ nm. Applying standard error propagation methods yields $\overline{V}^2=4\sigma_\Delta^2/\bar{\Delta}^2\simeq 0.14$. 

Second, to capture the broadening effects that arise from inelastic processes, we include a generic phenomenological dampening term in the self-energy of the form $\Sigma_{bath} = 2i\omega\gamma$. To keep the disorder discussion general, we postpone the analysis of the possible microscopic origin of $\gamma$ to a later section. For now, we note that $\gamma$ effectively encapsulates a range of energy-exchanging processes that couple to interlayer phase dynamics and lead to JPR damping. These may include incoherent tunneling of nodal and thermally excited quasiparticles, scattering by low-energy bosonic excitations such as optical phonons or spin fluctuations, and weak coupling to bound vortex–antivortex fluctuations among many others. Each of these mechanisms grows in relevance as temperature increases toward $T_c$, resulting in a progressive broadening and redshift of the JPR.
\vspace{0.5cm}

\textit{Nonlinear response and echo peak.---}
\begin{figure}
    \centering
    \includegraphics[width=0.95\linewidth]{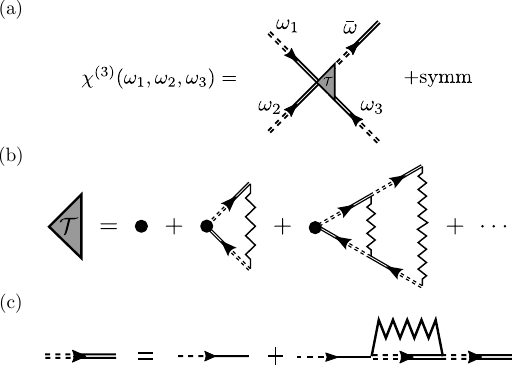}
    \caption{Sketch of the diagrammatic calculations, where the dot represents the bare interaction, and the zig-zag line is the Gaussian disorder. Single (Double) line propagators correspond to bare (dressed) retarded propagators. (a) Diagrammatic representation of $\chi^{(3)}$. (b) Disorder corrected vertex $\mathcal{T}$ within the nonpertubative noncrossing approximation. (c) Dyson equation for the propagators within the self-consistent Born approximation.}
    \label{fig: Diagrams}
\end{figure}
\begin{figure}
    \centering
    \includegraphics[width=\linewidth]{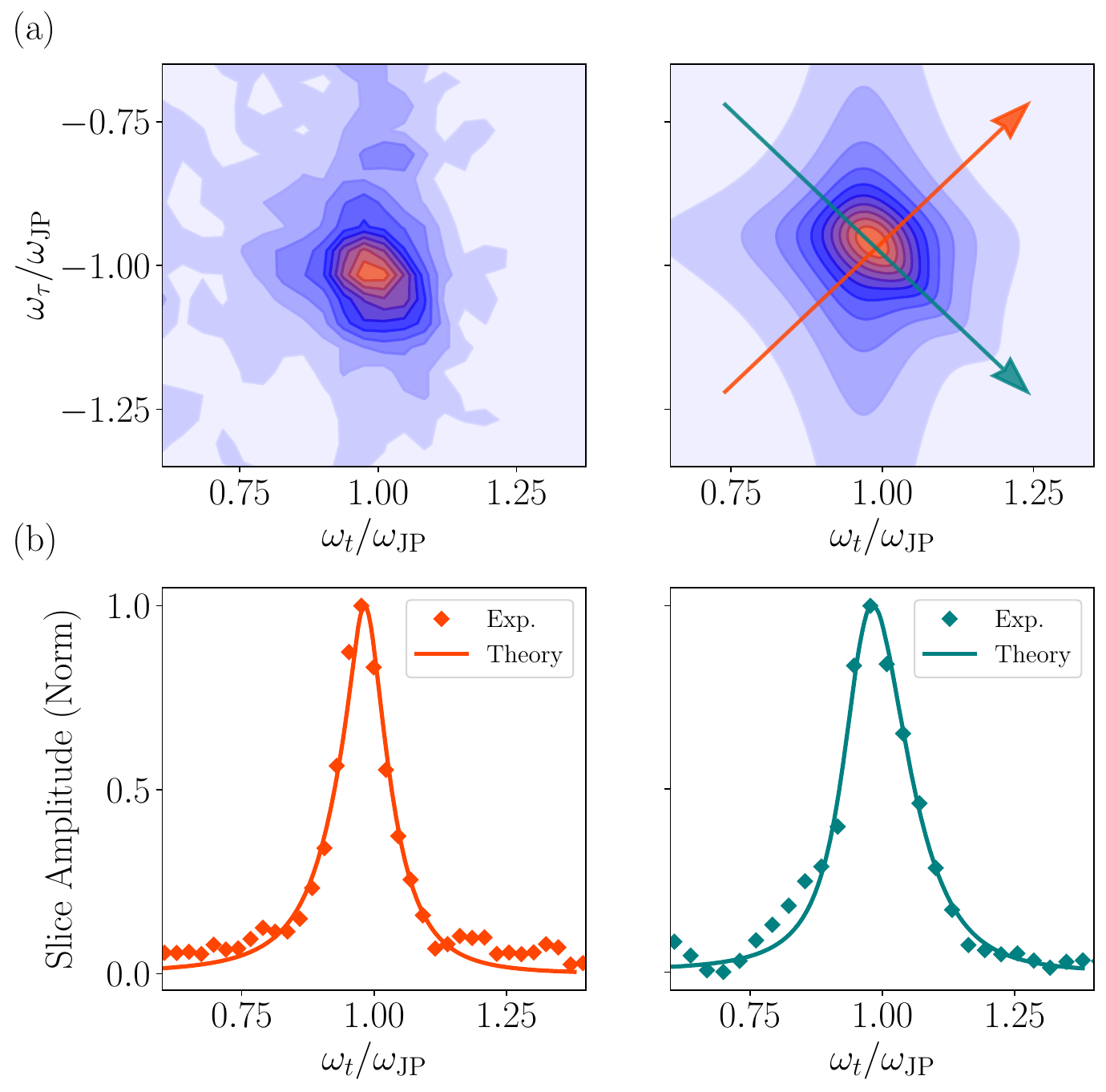}
    \caption{Comparison between the experimental and theoretical results for $T=15 K$. In (a) the experimental (left) and theoretical (right) echo peaks in the 2D map. In (b) the slices along the cross diagonal (left, orange) and diagonal (right, teal). The only fitting parameter is the inelastic scattering strength $\gamma_{\text{MB}}/\omega_{\text{JP}}\sim0.075$, while the elastic scattering strength is fully determined from STM data~\cite{kato2005inhomogeneous} and the microscopic parameters of the model.}
    \label{fig: Comparison}
\end{figure}
The calculation of $\chi^{(3)}$ for bosonic collective excitations is performed employing a non-crossing approximation for the disorder within the Keldysh path integral formalism~\cite{salvador2025Echo_theory}. Note that we assume all the external momenta to be zero due to the optical nature of the protocol. Furthermore, we restrict ourselves to the weak nonlinear regime discussed in~\cite{salvador2025Echo_theory}, where interaction induced quantum fluctuations are small and can be neglected. From a theoretical perspective, the central ingredient to capture the echo physics is the non perturbative vertex corrections induced by the static disorder. The schematic calculation of $\chi^{(3)}$ containing the disorder corrected vertex $\mathcal{T}$ is presented in Fig.~\ref{fig: Diagrams}(a) and (b). Note that the propagators in Fig.~\ref{fig: Diagrams} are dressed according to their corresponding self-consistent Dyson equation, as shown in Fig.~\ref{fig: Diagrams}(c). Within the weak nonlinear regime, $\mathcal{T}$ can be exactly expressed in terms of the self-energies that arise from both the elastic and inelastic processes~\cite{salvador2025Echo_theory}:
\begin{equation}
    \mathcal{T}(\omega_a,\omega_b) \sim \frac{2V_0^2\lambda(\boldsymbol{0};\omega_a,\omega_b)}{1-V_0^2\lambda(\boldsymbol{0};\omega_a,\omega_b)},
    \label{eqn: general_vertex}
\end{equation}
with
\begin{equation}
    \lambda(\boldsymbol{0};\omega_a,\omega
    _b) = \frac{1}{V_0^2}\frac{\Sigma_V(\omega_a)-\Sigma_V(\omega_b)}{\Sigma(\omega_a)-\Sigma(\omega_b)-2\omega_a^2+2\omega_b^2}.
\end{equation}
Here, $V_0^2 = 2\pi\omega_{\text{JP}}^4\overline{V}^2\xi^2$, $\Sigma(\omega) = \Sigma_V(\omega)+\Sigma_{bath}(\omega)$ is the total self-energy of the system in the perturbative nonlinear regime, and $\Sigma_V(\omega)$ is the is the Born self-energy given by
\begin{equation}
    \Sigma_V(\omega) = \frac{4V_0^2}{N L^2}\sum_{\boldsymbol{k},k_z}\mathcal{D}^R_{\boldsymbol{k},k_z}(\omega),
    \label{eqn: Born_self_energy}
\end{equation}
with $\left[\mathcal{D}^R_{\boldsymbol{k},k_z}(\omega)\right]^{-1} = 2\left(\omega^2-\omega_{\text{JP}}^2-L(k_z)\boldsymbol{k}^2-\Sigma(\omega)/2\right)$ the full retarded Green's function, and $L(k_z)$ the Fourier transform of $L_{nm}$, defined in Eq.~\eqref{eqn: def L}. In particular, for the echo diagonal $\omega=\omega_a=-\omega_b$, Eq.~\eqref{eqn: general_vertex} takes the simple form
\begin{equation}
    \mathcal{T}(\omega,-\omega) \sim \frac{\text{Im}\Sigma_V(\omega)}{\text{Im}\Sigma_{bath}(\omega)} = \frac{-\text{Im}\Sigma_V(\omega)/2}{\omega\gamma}.
    \label{eqn: vertex_self_energy_classical}
\end{equation}
Eq.~\eqref{eqn: vertex_self_energy_classical} completely characterizes the echo peak phenomenology as a function of the strength of  both elastic and inelastic scattering processes. In the inelastic limit ($\text{Im}\Sigma_V\ll\text{Im}\Sigma_{bath}$), vertex corrections due to the disorder are negligible, and the echo peak presents a symmetric shape, indicating the absence of disorder-induced broadening. Conversely, in the elastic limit ($\text{Im}\Sigma_V\gg\text{Im}\Sigma_{bath}$), vertex corrections due to the disorder dominate, and the vertex develops a sharp line of singularities along the echo diagonal for $\abs{\omega}>\omega_{\text{JP}}$. This pronounced feature along the echo diagonal gives rise to the characteristic almond-like shape, which heralds the presence of elastic scattering. Consequently, the echo peak uniquely disentangles the two broadening mechanisms, in contrast to linear-response measurements—such as reflectivity—which add the two contributions.

\begin{figure}
    \centering
    \includegraphics[width=\linewidth]{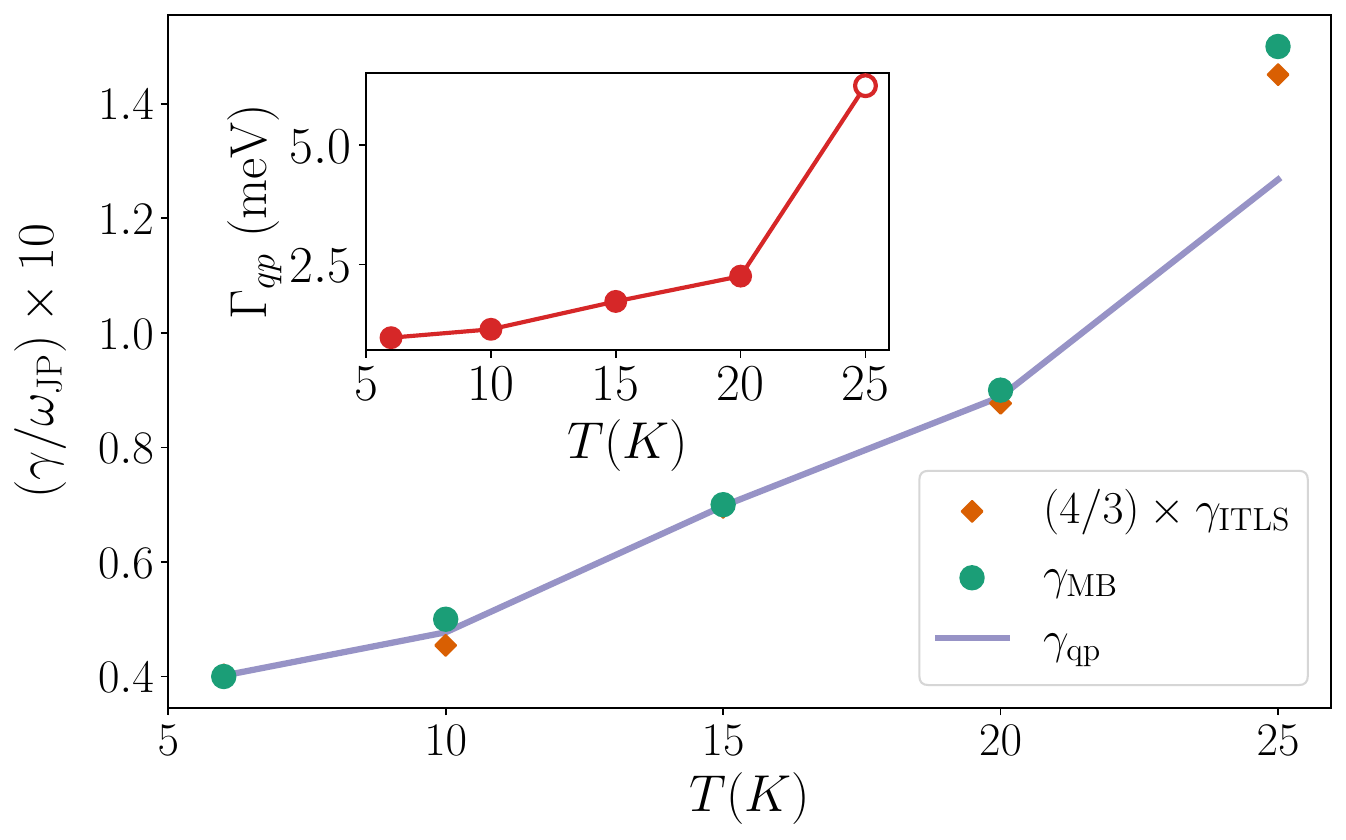}
    \caption{Comparison between the inelastic scattering fits from the independent two-level system (ITLS) forms, the theoretical fits employing the many-body (MB) framework, and the conductivity calculation. The inset shows the extracted electronic broadening. The last data point is shown as an open symbol, highlighting that fermionic quasiparticles alone cannot account for $\gamma_{\text{qp}}$ at 25 K, pointing to additional inelastic scattering mechanisms for the JPR.}
    \label{fig: gamma_comparison}
\end{figure}

We now proceed to evaluate $\chi^{(3)}$ for our model, produce the corresponding theoretical echo peaks, and compare them with the Josephson echoes in Ref.~\cite{Liu_2023_echo}. $\chi^{(3)}$ is obtained after evaluating Eq.~\eqref{eqn: Born_self_energy} employing $\xi$ and $\overline{V}^2$ extracted from~\cite{kato2005inhomogeneous}, quoted previously in the text, in combination with typical LSCO estimates of
$\epsilon_\infty=25$, $\lambda_{ab}=0.25\,\mu$m, and $s=6.6\,\text{\AA}$. We employ these values, together with the phenomenological fitting parameter $\gamma_{\text{MB}}$, which characterizes the strength on inelastic scattering processes, to fit the experimentally measured echo peaks in Ref.~\cite{Liu_2023_echo}. Within the temperature range of interest, the superfluid density is effectively temperature-independent~\cite{Kitano1999,Assa_2009_T_dependence}. Consequently, the elastic scattering rate is taken to be constant across all maps, consistent with the findings of Ref.~\cite{Liu_2023_echo}. We find excellent agreement between theory and experiments with $\gamma_{\text{MB}}/\omega_{\text{JP}}=\{ 0.04,0.05,0.07,0.09, 0.14 \}$ for the corresponding maps at temperatures $T=\{ 6,10,15,20,25 \}$K. A direct comparison with values extracted from standard two-level system fitting procedures $\gamma_{\text{ITLS}}$, as in Ref.~\cite{Liu_2023_echo}, may result in apparent discrepancies if the underlying many-body effects are not taken into account. Incorporating these effects, we find that a practical rule of thumb to relate the two is $\gamma_{\text{ITLS}} \simeq 0.75\,\gamma_{\text{MB}}$, see Fig.~\ref{fig: gamma_comparison} for an explicit comparison. This conversion applies for any linearly driven bosonic collective excitation; for a detailed discussion on the physical origin of the conversion factor see Ref.~\cite{Supplement}.

As a representative of this fitting procedure, we plot in Fig.~\ref{fig: Comparison} the comparison between the experimental and theoretical echo peaks (a), and their corresponding slices along the cross-diagonal and diagonal (b) for $T=15$ K. For completeness, we provide the fits for the other temperatures in the Supplemental material~\cite{Supplement}. We emphasize that no choice of $\gamma_{\text{MB}}$ can compensate for an inaccurate estimate of the disorder strength, given their distinct contributions to the asymmetry of the echo peak, as highlighted by Eq.~\eqref{eqn: vertex_self_energy_classical} and subsequent discussion. This analysis supports the conclusion that attributing the single-particle fermionic gap purely to superconductivity is consistent with the Josephson echo measurements, i.e., we find no indications of non-superconducting local orders that further contribute a gap to the electronic energy spectrum. This is in accordance to the conclusion reached in recent work based on shot noise STM measurements~\cite{niu2024equivalencepseudogappairingenergy}.
\vspace{0.5cm}

\textit{Origin of the inelastic scattering.---} The complexity of cuprates allows for a variety of elastic scattering mechanisms that may broaden the JPR. However, for the $6-25$K ($0.5-2.15$ meV) range spanned by the 2DTS experiments, and taking into account that $\omega_{\text{JP}}\sim8$ meV, we can discard some of them from energy considerations. We can safely disregard vortex-antivortex pair creation since both the JPR and the typical temperatures are energetically insufficient to create tightly bound vortex pairs with typical energies $\sim \pi \rho_s\sim 18$ meV, given a typical superfluid stiffness of $\rho_s\sim6$ meV (corresponding to $\lambda_{ab}=0.25\,\mu$m)~\cite{vonhoegen2025visualizing}. Similar reasoning excludes phonon and magnon induced broadening, with typical energies on the order 25 meV~\cite{phonon_LSCO} and 20 meV~\cite{Ikeuchi_2022_Spin_excitation} respectively. However, due to the $d$-wave symmetry of the superconducting gap in cuprates, gapless nodal quasiparticles are present at all temperatures. We argue that the low temperature inelastic scattering observed in 2DTS can be attributed to nodal fermionic quasiparticles and its thermal activation.

To characterize the effects of nodal quasiparticles on the JPR, we adapt the in-plane optical conductivity of Lee~\cite{Lee_1993_Localized_d_wave} and Hirschfeld et. al.~\cite{Hirschfeld_1994_microwave}, to the out-of-plane conductivity, such that
\begin{widetext}
\begin{equation}
    \text{Re}\,\sigma(\omega) = \frac{Me^2\expval{v_z^2}}{\pi L^2}\sum_{\boldsymbol{k}}{}^{'}\int d\Omega\,\frac{n_F(\Omega)-n_F(\omega+\Omega)}{\omega}\left[\text{Im}G(\Omega,\boldsymbol{k})\text{Im}G(\Omega+\omega,\boldsymbol{k})+\text{Im}F(\Omega,\boldsymbol{k})\text{Im}F(\Omega+\omega,\boldsymbol{k})\right],
\end{equation}
\end{widetext}
where $G$ and $F$ are respectively the dressed normal and anomalous components of the Nambu Green's function. Within the nodal approximation, we consider $E_{\boldsymbol{k}}\simeq\sqrt{v_F^2k_1^2+v_\Delta^2k_2^2}$, with $v_\Delta = 0.2\times10^5$ m/s and $v_F = 0.2\times10^5$ m/s. The prime summation indicates that it is only performed around one node, and $M=4$ is the number of nodes. Furthermore, we consider $\expval{v_z^2(k_z)}=(1/L)\sum_{k_z}v_z^2(k_z)=2s t_z^2$, with $t_z\simeq2$ meV the interlayer hopping strength. 
%
%
The effective inelastic scattering constant for the JPR is thus given by $\gamma_{\text{qp}}=\text{Re}\,\sigma(\omega)/\epsilon_0$.
We include a phenomenological broadening $\Gamma_{\rm{qp}}$ for the fermionic quasiparticles to fit the theoretical conductivity dampening extracted from the inelastic scattering of the echo peak.
We plot in Fig.~\ref{fig: gamma_comparison}, the comparison between the independent two-level system fitting forms by Siemens et al.~\cite{Siemens_2010}, the fits employing the many body formalism~\cite{salvador2025Echo_theory}, and the curve from the conductivity calculation as a function of temperature. We find good agreement for values of $\Gamma_{\rm{qp}}$ between 1 and 2.5 meV increasing with temperature, see the inset in Fig.~\ref{fig: gamma_comparison}. This range is compatible with the microwave measurements in $\mathrm{YBa_2Cu_3O_{7-\delta}}$~\cite{Hosseini_microwave,Turner_microwave}, which indicate a long lifetime, $\Gamma_{\rm qp}\sim 0.05$ meV, for nodal quasiparticles at temperatures well below $T_c$, and photoemission experiments in $\mathrm{Bi_2Sr_2CaCu_2O_{8+\delta}}$~\cite{Valla_1999_Arpes_BSSCO} and LSCO~\cite{Chang_ARPES_LSCO_2008}, which find elastic scattering rates on the order of 15 meV. 
Furthermore, LSCO is significantly more disordered than $\mathrm{YBa_2Cu_3O_{7-\delta}}$ and an order of magnitude increase in the inelastic scattering rate is to be expected. As a last remark, our microscopic calculation for $\gamma_{\text{qp}}$ is not enough to capture the last data point of Fig.~\ref{fig: gamma_comparison}, and therefore other inelastic scattering processes may be involved at 25 K beyond nodal quasiparticles.


\vspace{0.25cm}
\textit{Conclusions and outlook.---} In conclusion, we have demonstrated the capabilities of 2DTS on Josephson plasmons as a probe of the spatial inhomogeneity of the superfluid density. Owing to the unique capability of the echo peak to disentangle elastic and inelastic scattering processes, we concluded that the pairing gap is the sole responsible for the inhomogeneity in the fermionic energy gap observed in tunneling experiments. Consequently, from a 2DTS perspective, we find no indication of the presence of competing non-superconducting local orders with a fermionic gap. We have characterized the low temperature dependence of the inelastic scattering processes and attributed it to the presence of nodal quasiparticles.

Although our analysis has focused on Josephson plasmons in high-Tc cuprates, the probing scheme we introduce is broadly applicable. Harnessing the nonlinear response of collective excitations tied to local electronic orders provides a powerful means to reveal both the nature and the disorder of the underlying electronic landscape. Looking ahead, we anticipate this approach will open new avenues for investigating competing and intertwined orders across a wide range of strongly correlated quantum materials, establishing multidimensional spectroscopy as a versatile tool for diagnosing emergent many-body phenomena.

We thank L. Benfatto, P. A. Lee, F. Marijanovic for insightful discussions. A.G.S., I.M., and E.D. acknowledge the Swiss National Science Foundation (project $200021\_212899$), ETH-C-06 21-2 equilibrium Grant with project number 1-008831-001, and  NCCR SPIN, a National Centre of Competence in Research, funded by the Swiss National Science Foundation (grant number 225153) for funding. A.L. was supported by the U.S. Department of Energy, Office of Basic Energy Sciences, under Contract No. DE-SC0012704.

\vspace{1cm}

%


\appendix
\newpage
\setcounter{figure}{0}
\setcounter{equation}{0}

\renewcommand{\thepage}{S\arabic{page}} 
\renewcommand{\thesection}{S\arabic{section}} 
\renewcommand{\thetable}{S\arabic{table}}  
\renewcommand{\thefigure}{S\arabic{figure}} 
\renewcommand{\theequation}{S\arabic{equation}} 

\onecolumngrid

\begin{center}
\textbf{\Large{\large{Supplemental Material \\ \vspace{0.75cm}Gap Inhomogeneity in Cuprates: a view from Two-Dimensional Josephson Echo Spectroscopy}}}
\end{center}

\section{Origin of the 4/3 factor between the independent two-level system fit form and the Many-body theory.}

The apparent discrepancy between the homogeneous broadening extracted from the independent two-level system (ITLS) fitting forms $\gamma'$~\cite{Siemens_2010} and the Many-body theory used here $\gamma$~\cite{salvador2025Echo_theory} can be resolved by investigating the characteristic oscillations of each model. While the two-level system decays in time following $\exp{(-\gamma' t)}$, a single bosonic collective in the presence of a bath obeys $\exp{(-\gamma t/2)}$. This is due to the fact that
\begin{equation}
    \mathcal{D}^R(\omega,\boldsymbol{k}) \sim \frac{1}{\omega^2-\epsilon_k^2+i\omega\gamma}\approx \frac{1}{(\omega+i\gamma/2)^2-\epsilon_k^2} = \frac{1}{2\epsilon_k}\left(\frac{1}{\omega+i\gamma/2-\epsilon_k} - \frac{1}{\omega+i\gamma/2+\epsilon_k}\right).
\end{equation}
As discussed in the main text of Ref.~\cite{salvador2025Echo_theory}, there exists a second family of processes that need to be considered when dealing with collective bosonic excitations. These are processes involving the coexistence of 3 bosonic excitations in the system during the second time delay. In processes involving single excitations, the decay goes as $\exp[-\gamma (\tau+t)/2]$, while the ones involving 3 excitations during the second time decay is given by $\exp[-\gamma (\tau+3t)/2]$. Since the ITLS fit only provides a single $\gamma'$, it effectively sees an average of both processes which contribute with equal strength to the nonlinear response. Thus, the effective gamma is obtained by averaging over all time delays: $\gamma'= \frac{1/2+1/2+1/2+3/2}{4}\gamma = \frac{3}{4}\gamma$.

\section{Experimental and Theory comparison of the echo signature for all Temperatures}

\begin{figure}
    \centering
    \includegraphics[width=0.85\linewidth]{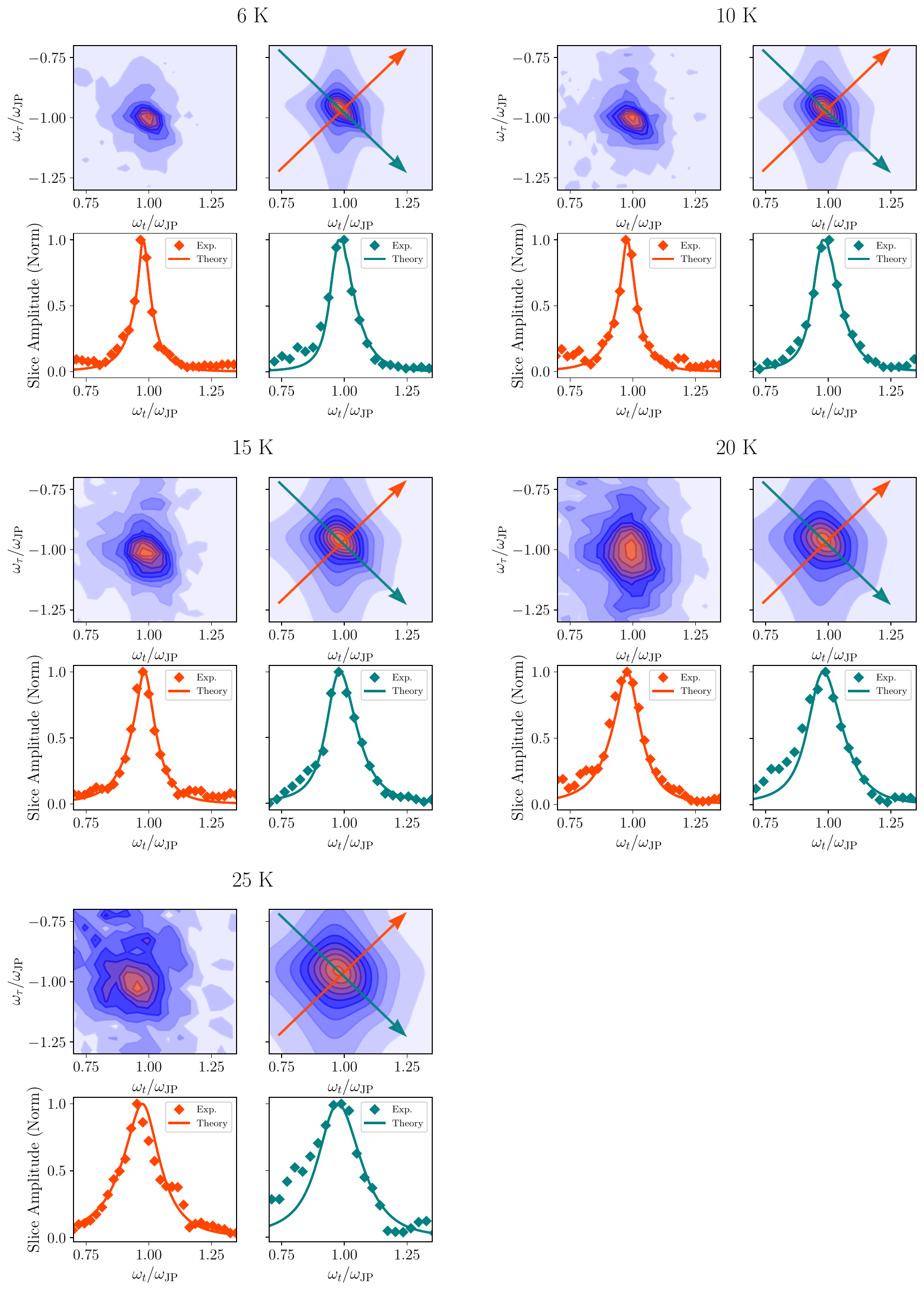}
    \caption{Comparison between the experimental and theoretical results for all temperatures. In the top left (right) of each sub-panel the experimental (theory) echo peaks in the 2D map. In the bottom row, the slices along the cross diagonal (left, orange) and diagonal (right, teal).}
    \label{fig: Comparison_all_temp}
\end{figure}

In this section we present the equivalent of Figure 2 for all temperatures, where the two-dimensional THz spectroscopy protocol was carried over. The comparison for a total of 5 temperatures, ranging from 6K to 25K, are presented in Fig.~\ref{fig: Comparison_all_temp}. We want to further reiterate that the strength of the disorder is fixed via the STM measurements~\cite{kato2005inhomogeneous} throughout the temperature study.

\end{document}